\documentclass{article}
\usepackage[preprint]{spconf}
\usepackage{amsmath,graphicx,listings,color, setspace, balance}

\title{Executing Dynamic Data Rate Actor Networks on OpenCL Platforms}
\name{J. Boutellier, I. Hautala}
\address{Center for Machine Vision and Signal Analysis\\
	University of Oulu, Finland}
\begin{document}

\copyrightnotice{\copyright\ IEEE 2016}
                 \toappear{Accepted to {\it Proc.\ IEEE SiPS 2016,
                    Oct 26-28, 2016, Dallas, Texas}}

\ninept
\maketitle
\begin{abstract}
Heterogeneous computing platforms consisting of general purpose processors (GPPs) and graphics processing units (GPUs) have become commonplace in personal mobile devices and embedded systems. For years, programming of these platforms was very tedious and simultaneous use of all available GPP and GPU resources required low-level programming to ensure efficient synchronization and data transfer between processors. However, in the last few years several high-level programming frameworks have emerged, which enable programmers to describe applications by means of abstractions such as dataflow or Kahn process networks and leave parallel execution, data transfer and synchronization to be handled by the framework.

Unfortunately, even the most advanced high-level programming frameworks have had shortcomings that limit their applicability to certain classes of applications. This paper presents a new, dataflow-flavored programming framework targeting heterogeneous platforms, and differs from previous approaches by allowing GPU-mapped actors to have data dependent consumption of inputs / production of outputs. Such flexibility is essential for configurable and adaptive applications that are becoming increasingly common in signal processing. \textit{In our experiments it is shown that this feature allows up to 5$\times$ increase in application throughput.}

The proposed framework is validated by application examples from the video processing and wireless communications domains. In the experiments the framework is compared to a well-known reference framework and it is shown that the proposed framework enables both a higher degree of flexibility and better throughput.

\end{abstract}
\section{Introduction}
\label{sec:intro}

Programming of graphics processing units (GPUs) found on heterogeneous computing platforms has required the use of OpenCL or Cuda until the last few years. Even though the basic usage of these languages can be considered rather straightforward, tapping the full computational potential of the platform, including all general purpose processors (GPPs) and the GPU simultaneously, is a very complex task that requires specialized expertise.

To this end, the research community has invested considerable effort in the development of programming frameworks \cite{Hyunh14, Schor13, Sbirlea12} that would relieve the programmer from the task of writing low-level code for optimized data transfers between the GPPs and GPU and valid synchronization between computations. Existing frameworks have found Kahn Process networks \cite{Kahn74} or dataflow abstractions \cite{Lee87} to be suitable for simplifying the programming effort.

Unfortunately, even the most advanced programming frameworks have restrictions that limit their applicability to a certain class \cite{Schor13} of (signal processing) applications, or fail to provide significant performance advantage \cite{Lund15} when compared to manually written OpenCL or Cuda programs. This paper addresses one of the fundamental limitations of existing solutions by introducing a novel dataflow-flavored programming framework that allows executing dynamic data rate applications on OpenCL / GPU devices. In the experiments it is shown that this feature allows up to a 5$\times$ increase in application throughput.

In detail, the proposed framework features:
\begin{itemize}
  \item An abstraction for expressing applications as a network of dataflow actors,
  \item Concurrent execution of GPP and GPU mapped actors,
  \item Support for executing dataflow actors with dynamic data rates, also on the GPU,
  \item Support for applications with initial (delay) tokens.
\end{itemize}

The functionality of the framework is demonstrated by benchmarking two applications, video motion detection and dynamic predistortion filtering, and by comparing the results to the well-known DAL framework \cite{Schor12}.

The rest of this paper is organized as follows:
Section~\ref{sec:background} introduces the dataflow abstraction and presents related work, Section~\ref{sec:proposed} details the central contributions of this work, Section~\ref{sec:experiments} presents experimental evaluation of the proposed approach, Section~\ref{sec:discussion} discusses the results, and Section~\ref{sec:conclusion} concludes the paper.

\section{Background}
\label{sec:background}

\subsection{Dataflow Abstraction}
\label{ssec:rvccal}

\begin{figure}
\centering
\includegraphics[width=0.88\linewidth]{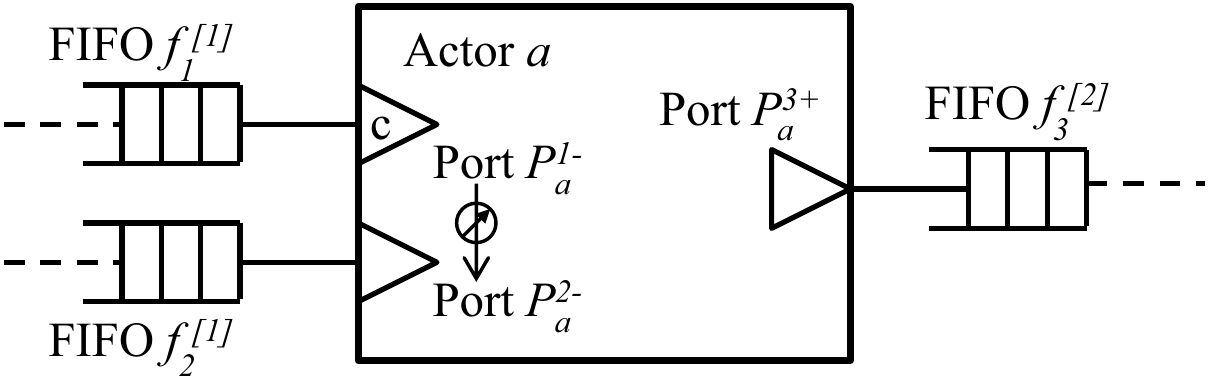}
\caption{Dataflow actor $a$ that has two input ports and one output port.}
\label{fig:actor}
\end{figure}

In the dataflow abstraction \cite{Lee87}, applications are composed of \textit{actors} that perform computations on data that is quantized into \textit{tokens}. Each actor is created when the application is launched, and is terminated when the whole application has finished. Actors acquire tokens from their input \textit{ports} and produce computation results to their output ports. Token communication between actors is handled by channels that have an order-preserving FIFO (First-In-First-Out) behavior. An actor (See Fig.~\ref{fig:actor}) performs a computation by \textit{firing}, which can include consuming tokens from input ports at the beginning of the firing, and producing tokens to the actor output ports at the end of the firing. A central feature of the dataflow abstraction is that computations are triggered by the availability of data, in contrast to, for example, time-triggered abstractions \cite{Henzinger01}.

In literature, a wide variety of dataflow Models of Computation (MoC) have been presented. One of the most important factors that differentiate a dataflow MoC from another concerns the token communication \textit{rates} when an actor reads from, or writes to, a channel to which it connects. In this sense, the most restricted dataflow MoC is \textit{homogeneous synchronous dataflow} (HSDF) \cite{Lee87}, where an actor must read exactly one token from each of its input ports and produce exactly one token to each of its output ports on each firing. \textit{Synchronous dataflow} (SDF) \cite{Lee87} is slightly more expressive as it allows token rates larger than one, as is \textit{cycle-static dataflow} (CSDF) \cite{Bilsen96} that goes beyond SDF by allowing tokens rates to vary in repetitive cycles.

The aforementioned MoCs (HSDF, SDF, CSDF) are restricted in the sense that they disallow \textit{data dependent} changes to the token rates, which is a required feature as, for example, video decoders \cite{Mattavelli10} and Software Defined Radio applications \cite{Berg08} introduce behavior that cannot be captured by static data rates. To achieve this, \textit{dynamic dataflow} MoCs are required. Examples of dynamic dataflow MoCs are \textit{boolean dataflow} (BDF) \cite{Buck93}, \textit{enable-invoke dataflow} (EIDF) \cite{Plishker08} and \textit{dataflow process networks} (DPN) \cite{Lee95} that allow port token rates to change at application run time. Some formulations \cite{Lee09, Tretter15} also allow interpreting Kahn process networks (KPN) \cite{Kahn74} as a kind of a dynamic dataflow MoC.

The proposed framework is based on dynamic dataflow, and the essential features of its computation model are presented next, following the notation adopted from \cite{Boutellier15T}.

\subsection{Dataflow Model of the Proposed Framework}
\label{ssec:ourmodel}
In the proposed framework an application is described as a network $\aleph = (A, F)$, where $A$ is a set of actors and $F$ is a set of FIFO communication channels that interconnect the actors. Each actor $a (\in A)$ may have 0 or more input ports $P_a^{-}$ and zero or more output ports $P_a^{+}$. If an actor $a$ has zero input ports it is called a \textit{source actor}, and if it has zero output ports it is called a \textit{sink actor}.

Each FIFO channel $f \in F$ has an associated token rate $f^{[r]}$, where $r$ is a positive integer. If actor $a$ is connected to FIFO channel $f_k$ (where $k$ is the index number of the FIFO) through its output port $P_a^{k+}$, the output port adopts the token rate $r$ of the FIFO buffer. Same applies for input ports. Both reads and writes to channels are \textit{blocking}, such that the execution of the reading (writing) actor stalls until sufficient tokens are (space is) available.

An actor $a$ may be \textit{static} or \textit{dynamic}. Dynamic actors have one \textit{control input port} and \textit{regular ports}. A control port always has a fixed token rate of 1 (and hence, the FIFO to which it connects, must also have a token rate of 1), whereas a regular port of a dynamic actor may have two token rates: 0 and $r$, where $r$ is the token rate of the FIFO channel $f_k^{[r]}$ to which $P_a^k$ connects. For static actors, on the other hand, each input $P_a^{k-}$ and output $P_a^{k+}$ port of $a$ always has a single fixed token rate inherited from the FIFO $f_k$ to which it connects.

Figure~\ref{fig:actor} depicts an example of this. Dynamic actor $a$ is connected to three FIFO channels: $f_1$, $f_2$ and $f_3$ through its ports $P_a^{1-}$, $P_a^{2-}$ and $P_a^{3+}$. FIFOs $f_1$, $f_2$ and $f_3$ have token rates of 1, 1, and 2, respectively. Port $P_a^{1-}$ is the control port of actor $a$, denoted with a "c" in the figure. Values of the tokens consumed from the control port set the token rate of input port $P_a^{2-}$ to either 0 or 1, however the mapping of control token port values to token rates of $P_a^{2-}$ is left unspecified here.

When the dynamic actor $a$ fires, it first consumes one token from its control port. Then, based on the token value of the consumed control port token, the token rate of each regular input port and regular output port $P$ of $a$ is fixed to either 0 or $r$ ($r$ being adopted from the associated FIFO $f^{[r]}$) \textit{for the duration of this firing}. After fixing the token rates, actor $a$ consumes tokens from each input port $P_a^{-}$ that has a non-zero token rate for the duration of this firing. Based on the tokens consumed from the input ports, $a$ performs computations and finally produces tokens to each output port $P_a^{+}$ that has a non-zero token rate.

Any FIFO buffer $f \in F$ that is \textit{not} connected to a control port is allowed to have 0 or 1 \textit{initial tokens} (delays) irrespective of the token rate $f^{[r]}$. Initial tokens are data that is present in FIFO buffers before any actor  $a \in A$ has fired, and is normally used to model feedbacks in signal processing systems, for example, in Infinite Impulse Response (IIR) filters.

It is necessary to point out that the computation model described above cannot be reduced to single-rate dataflow (HSDF) due to the allowed presence of initial tokens. Also, the described model bears resemblance to boolean dataflow \cite{Buck93}, however a more detailed analysis of the similarities and differences must be presented elsewhere due to limitations in presentation space.

\subsection{Related Programming Frameworks}
\label{ssec:relatedwork}

A number of programming frameworks targeting heterogeneous platforms have emerged in the last few years. The frameworks described in \cite{Boulos14}, \cite{Sbirlea12} and \cite{Gautier13} represent task-based programming approaches, where tasks are spawned, executed and finished, and their interdependencies are expressed as a directed acyclic graph. The proposed approach, in contrast, is based on actors that are created once at initialization and run as independent entities, communicating with each other until termination of the application.

A recent article \cite{Hyunh14} presents a framework that enables deploying applications written in the StreamIt language \cite{Thies02} to GPUs. Compared to this work, the significant difference is that the StreamIt language heeds the SDF MoC that does not allow data dependent execution paths or dynamic data rates, whereas the proposed framework allows dynamic data rates as described above. The same restriction of dynamic data rates also applies to two recent works \cite{Lund15, Boutellier15S} that discuss deployment of RVC-CAL dataflow programs to heterogeneous architectures.

The DAL framework \cite{Schor12} is based on Kahn process networks and also has an extension \cite{Schor13} for targeting heterogeneous systems with OpenCL enabled devices. In terms of OpenCL / GPU acceleration this framework is limited to the SDF MoC that disallows dynamic data rates.

\section{The Proposed Framework}
\label{sec:proposed}

This section provides an overview of the proposed framework: Subsection~\ref{ssec:actors} explains how the programmer expresses actors for the proposed framework, Subsection~\ref{ssec:channels} describes the proposed inter-actor communications techniques, Subsection~\ref{ssec:concurrency} details the implementation of concurrency and finally Subsection~\ref{ssec:toolchain} gives an overview of the concrete framework implementation.

\subsection{Description of Actors}
\label{ssec:actors}

In the proposed framework each actor consists of the mandatory \textit{fire} function, and optional \textit{init}, \textit{control}, and \textit{finish} functions. The \textit{fire} function describes the actor's behavior upon firing and comprises the reading of regular input ports, computation and writing to regular output ports. The optional \textit{init} and \textit{finish} functions are only executed once on application initialization and termination, and are mainly useful for source and sink actors to start and end interfacing with I/O. The \textit{control} function is only required for dynamic actors and is executed once for each firing of the actor, right before invoking the \textit{fire} function. The \textit{control} function is responsible for setting the data rates of regular ports.

At design time the programmer chooses whether the actor is going to be executed on an OpenCL / GPU device or on one of the general purpose cores. Depending on the choice, the actor functionality is written either in OpenCL C, or in the conventional C language. The proposed framework provides a minimal API that essentially provides functions for inter-actor communication, such as \texttt{fifoWriteStart}, \texttt{fifoWriteEnd}, etc.

This formulation, where actors consist of \textit{init}, \textit{fire}, and \textit{finish} functions is identical to the DAL \cite{Schor12} framework. However, the \textit{control} function, especially required for enabling dynamic data rate actors on OpenCL / GPU devices, is specific to this framework. The \textit{control} function takes one control token as its input and is required to set the data rate (to 0 or $r$ as defined in Subsection~\ref{ssec:ourmodel}) of each regular input and regular output port for the duration of one firing. The proposed framework does not impose limitations regarding the mapping of control token values to the token rates of ports.


\begin{figure}
\centering
\includegraphics[width=\linewidth]{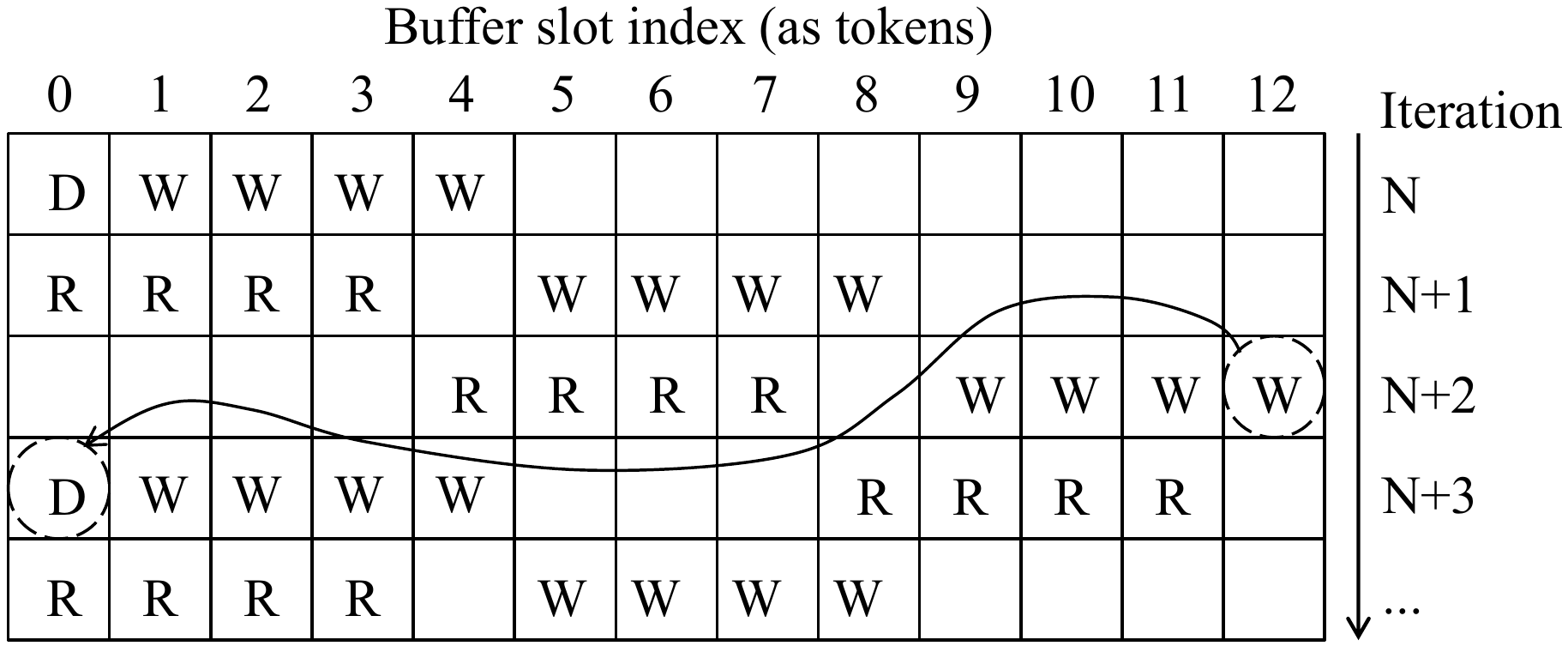}
\caption{Buffer access pattern in the case of a delay token for token rate 4.}
\label{fig:triplebuffer}
\end{figure}

\subsection{Communication Channels}
\label{ssec:channels}

A communication channel in the proposed framework connects exactly one output port of an actor to exactly one input port of another actor, heeding FIFO behavior. In contrast to other (e.g. \cite{Schor12}) programming frameworks, the token capacity of a communication channel $f$ cannot be arbitrarily chosen by the programmer, but is exactly specified as
\begin{equation}
C_f = 
    \begin{cases}
	S_f * (r * 3 + 1), & \text{if $f$ has a delay token} \\
	S_f * (r * 2), & \text{otherwise,} \\
    \end{cases}
\end{equation}
where $r$ is the token rate, and $S_f$ is the size (e.g. in bytes) of one token of FIFO $f$.

As a communication channel $f$ assumes to receive $r$ new tokens on each write, and to output $r$ tokens on each read, we see that a regular channel (the \textit{otherwise} case in Eq. 1) is essentially a double buffer that allows simultaneous reading and writing to the channel. However, for channels that contain an initial (delay) token, the channel is implemented as a slightly more complex buffer that implements a specific access pattern to enable simultaneous reads and writes to the channel. This is depicted in Fig.~\ref{fig:triplebuffer} with an example case of $r = 4$. 

At application initialization the initial token in the channel, displayed with $D$ in Fig.~\ref{fig:triplebuffer} resides in buffer slot 0. The first write to the channel occupies slots 1 ... 4, whereas the first read consumes tokens from slots 0 ... 3 and so forth. The third write to the channel reaches the end (slot 12) of the buffer, followed by an explicit data copy from slot 12 to slot 0, and the access pattern starts to repeat. It is important to note that the access pattern is repetitive and can be generalized to any token rate $r \in [1,\inf[$. 

Looking at Fig.~\ref{fig:triplebuffer}, it is evident that this solution is not minimal in terms of memory footprint, but it was chosen as it offers 1) uncompromized throughput and 2) transparency to the application programmer. Conventional ring buffers were considered inadequate, as OpenCL / GPUs offer the best combination of performance and ease of programming when input and output data to kernels is provided as contiguous arrays. 

The memory footprint overhead of this solution is slightly more than 50\% (depending on the token rate) when compared to regular double buffers. Essentially the same triple-buffer solution can also be generalized to 2 or more delay tokens, however due to limits in presentation space the generalization is omitted here.


\subsection{Concurrency, Scheduling and Actor-to-Core Mapping}
\label{ssec:concurrency}

The proposed framework has been designed to enable maximally parallel operation. Parallelism is based on threading, such that each actor runs on an operating system (OS) thread of its own, regardless whether the actor is targeted to OpenCL / GPU devices or to one of the general purpose cores. Each actor thread is created once at application startup, and is canceled after the application has terminated. Similar to the DAL framework \cite{Schor12, Schor13}, synchronization of data exchange over FIFO channels is based on \textit{mutex} locks and blocking communication: if an actor attempts to read a channel that has less tokens than the actor requires, the reading actor blocks until sufficient data is available. On one hand, this enables very efficient multiprocessing, but on the other hand makes the MoC somewhat more restricted than e.g. that of DPNs \cite{Tretter15}.

As each and every actor is instantiated as a separate thread using the GNU/Linux \textit{pthreads} library, the scheduling of actor firings (heeding data availability) is left to the OS. If the programmer so chooses, the framework allows fixing of actors to specific GPP cores, otherwise the OS chooses the core on which an actor firing is executed.

It is necessary to state that alternatively to the adopted OS threading based concurrency, it would also have been possible to build concurrency and synchronization on top of OpenCL events, however this would have limited the applicability to platforms where both the GPPs and GPUs have OpenCL drivers. The adopted OS threading based solution, however, is beneficial due to its backwards compatibility: with this solution it is possible to jointly synchronize and run also non-OpenCL compatible GPPs with GPUs.

\begin{figure}
\centering
\includegraphics[width=\linewidth]{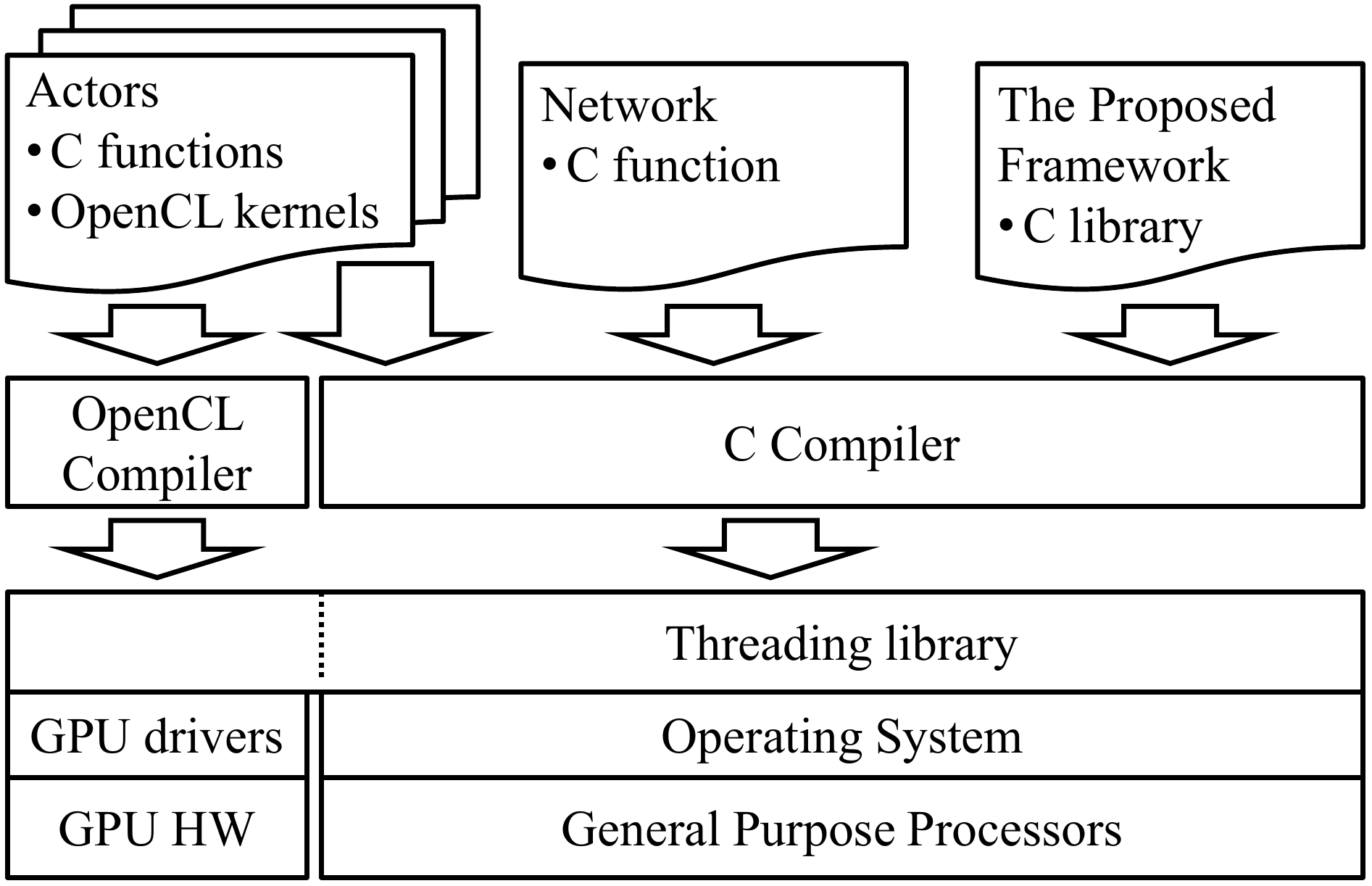}
\caption{Overview of the proposed framework.}
\label{fig:framework}
\end{figure}

\subsection{Overview of the Design Flow}
\label{ssec:toolchain}

Fig.~\ref{fig:framework} depicts a hierarchical view of the proposed framework. The implementation of the framework is written as a C library with an API that is called by the actors and by the actor network description. The actors that constitute the application are expressed as C functions, or as OpenCL kernels for OpenCL compatible targets, whereas the actor network is defined with a C function.

\begin{table}
\caption{Memory allocated to communication buffers in Megabytes.}
\label{table:buffers}
\begin{tabular}{p{3.0cm}p{0.7cm}p{0.7cm}p{1.0cm}p{1.0cm}}
\hline\noalign{\smallskip}
Framework & DAL & Prop. & DAL & Prop.\\
\textit{target} & MC & MC & Heterog. & Heterog. \\
\hline 
Motion Detection  & 0.77 & 0.85 & 3.69 & 3.46 \\
\hline
Dynamic Predistortion  & 11.5 & 11.5 & \textit{n/a} & 11.5 \\
\hline 
\noalign{\smallskip}
\end{tabular}
\end{table}

\begin{table*}
\caption{Platforms used for experiments.}
\label{table:platforms}
\begin{tabular}{p{0.8cm}p{5.5cm}p{6.6cm}p{3.2cm}}
\hline\noalign{\smallskip}
Tag & GPPs & GPU & Operating System\\
\hline 
Carrizo & AMD Pro A12-8800B (2.1 GHz, 4 cores) & AMD Radeon R7, OpenCL 2.0, driver 15.30.3 & Ubuntu 14.04, g++ 4.8.4 \\
\hline 
i7 & Intel Core i7-4770 (3.4 GHz, 4 cores) & AMD Radeon HD 7750, OpenCL 1.2, driver 15.20.3 & Ubuntu 14.04, g++ 4.8.4 \\
\hline 
\noalign{\smallskip}
\end{tabular}
\end{table*}

\begin{figure}
\centering
\includegraphics[width=\linewidth]{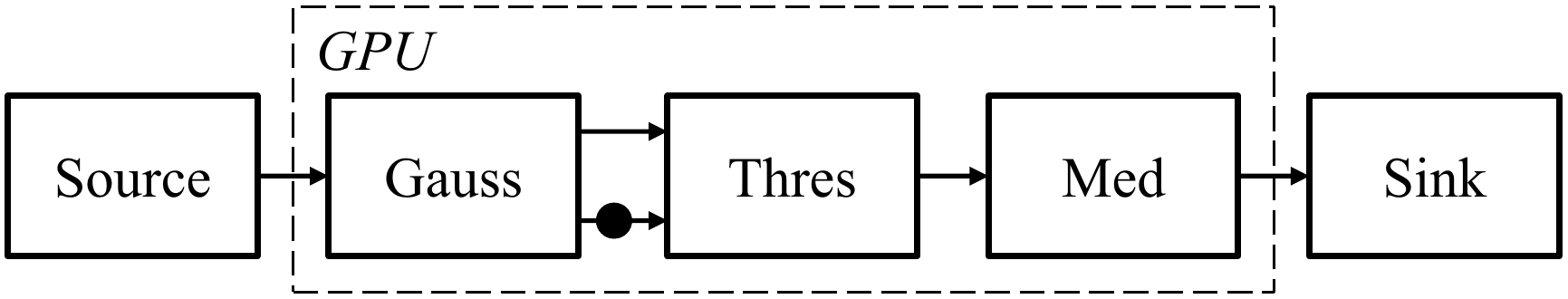}
\caption{The Motion Detection application.}
\label{fig:video}
\end{figure}

\begin{figure}
\centering
\includegraphics[width=0.76\linewidth]{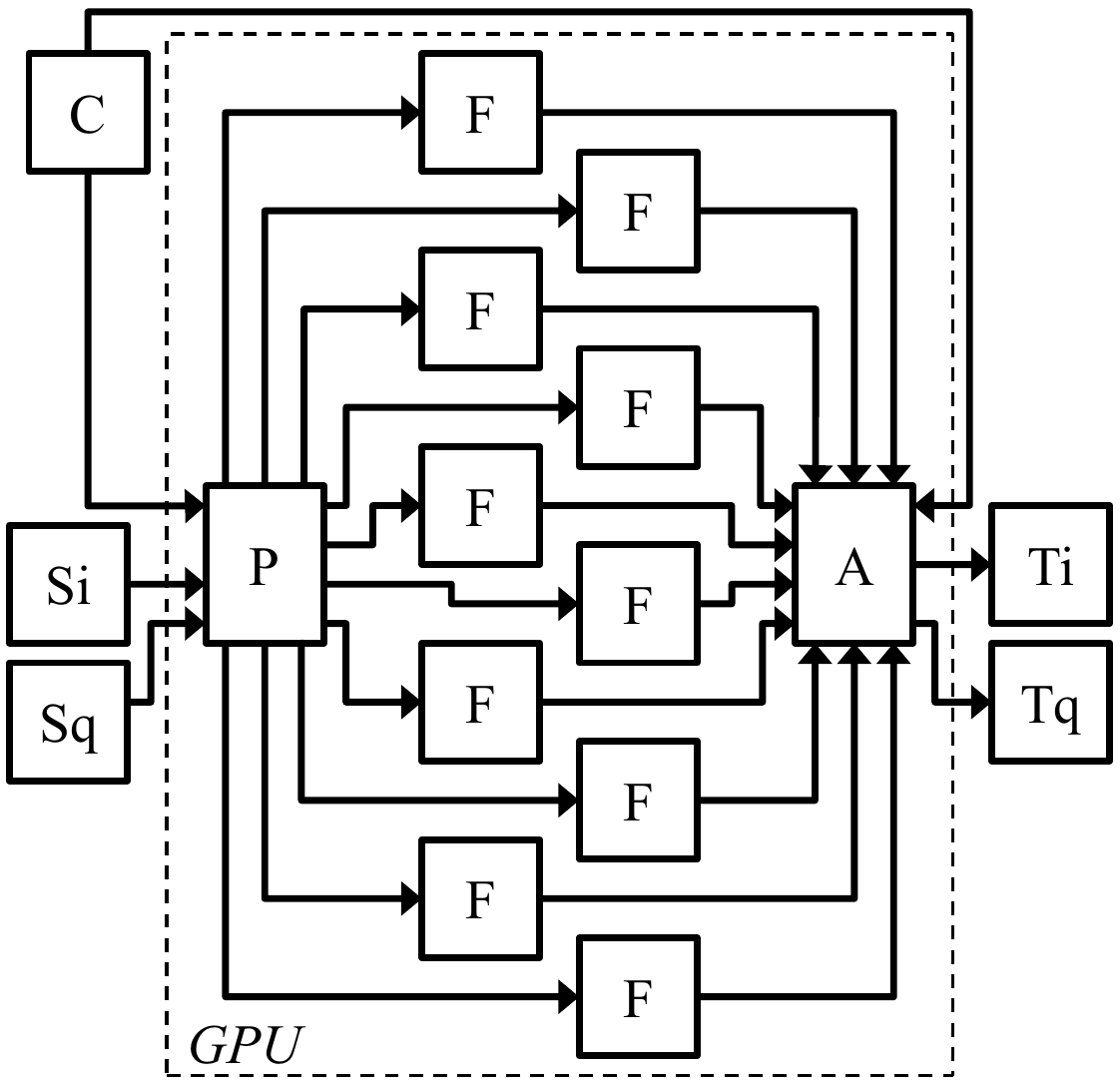}
\caption{The Dynamic Predistortion application.}
\label{fig:dpd}
\end{figure}

\section{Experiments}
\label{sec:experiments}

To validate the functionality and performance of the proposed framework, two applications were benchmarked on two different heterogeneous platforms that are described in Table~\ref{table:platforms}. The Carrizo chip features an integrated graphics processor, a solution that minimizes the data transfer times between the GPU and the GPP cores. The other platform, i7, represents a traditional solution where GPP cores communicate with the GPU over a PCI Express bus and thus the data transfer times between the GPP and GPU are non-negligible.

The DAL framework \cite{Schor12, Schor13} was used as a reference, and the code of the two applications was adapted from the proposed framework to DAL with minimal required changes. The DAL platform only allows fixed actor-to-core mappings, whereas the proposed framework also allows letting the OS to select the best core for execution, an option called \textit{free mapping} in the results.

\subsection{Video Motion Detection}

The first application used in our experiments is 8-bit grayscale video Motion Detection that consists of five actors, as shown in Fig.~\ref{fig:video}. The source and sink actors are always executed on GPP cores and are essentially responsible for reading and writing data from/to mass storage. The Gauss actor performs 5$\times$5 pixels Gaussian filtering on the input data, followed by the Thres actor that subtracts consecutive frames and performs pixel thresholding against a fixed constant value. To avoid exceeding frame boundaries the Gauss actor skips filtering for two pixel rows in the frame top and frame bottom. Finally, the Med actor performs 5-pixel median filtering to reduce noise from the generated motion map. One of the communication channels between the Gauss and Thres actors bears a dot in Fig.~\ref{fig:video} and depicts an initial token. The initial token is a one-frame delay that enables the functionality of consecutive frame subtraction.

To enable comparison with previous works, the frame size used was 320$\times$240, which resulted in the token size being 76800 bytes. In GPP-only execution the token rate on all channels was kept at one, as increasing the token rate did not have a measurable performance effect. GPU acceleration was applied to Motion Detection by mapping the Gauss, Thres and Med actors to the GPU. Total amount of memory used for buffers is shown in Table~\ref{table:buffers} for each configuration. The Motion Detection application is essentially the same as the one used in our previous work \cite{Boutellier15G}, however re-written for the proposed framework. In \cite{Boutellier15G} the functionality of Thres and Med actors was implemented in a single actor.

\subsection{Dynamic Predistortion Filtering}

Dynamic Predistortion (DPD) filtering (Fig.~\ref{fig:dpd}) was used as the second application test case. The algorithm is used in wireless communications to mitigate transceiver impairments, and to a great extent consists of parallel 10-tap FIR filters. Functionally, the filter is identical to the one presented in our previous work \cite{Boutellier15G}, but the actor descriptions have been rewritten for the proposed framework.

DPD significantly differs from the Motion Detection application in the sense that it features actors with dynamic data rates: Fig.~\ref{fig:dpd} shows the configuration (C) actor that at run-time periodically reconfigures the Poly (P) and Adder (A) actors to select which set of the FIR filters is used to process the input signal. The reconfiguration period was set to once every 65536 samples, and the number of active filter actors is allowed to change arbitrarily between 2 and 10. The run time reconfiguration used here is defined by an external input and cannot be modeled e.g. by the CSDF MoC.

The DPD application computes on complex floating point numbers, which were represented as a pair of single precision floats. To this end, all edges in Fig.~\ref{fig:dpd} \textit{inside} the "GPU" box represent a pair of edges, one for the real part and one for the imaginary part. Hence, the total number of FIFO channels is 46 in this application.

\begin{table}
\caption{Throughput for Motion Detection in \textit{frames per second} on multicore (MC) and MC+GPU (Heterog.) targets. $^\dagger$ see Subsect.~\ref{ssec:results}}
\label{table:motion}
\begin{tabular}{p{1.2cm}p{0.7cm}p{0.7cm}p{0.7cm}p{1.3cm}p{1.3cm}}
\hline\noalign{\smallskip}
Framework & DAL & Prop. & Prop. & DAL & Prop.\\
\textit{target} & MC & MC & MC & Heterog. & Heterog. \\
\textit{mapping} & fixed & fixed & free & - & - \\
\hline 
Carrizo  & 400 & 485 & 486 & 2915$^\dagger$ & 4614 \\
\hline 
i7 & 872 & 1138 & 1135 & 4320$^\dagger$ & 6063 \\
\hline
\noalign{\smallskip}
\end{tabular}
\end{table}

\begin{table} [t]
\caption{Throughput for Dynamic Predistortion in \textit{Megasamples per second} on multicore (MC) and multicore+GPU (Heterog.) targets.}
\label{table:dpd}
\begin{tabular}{p{1.2cm}p{0.7cm}p{0.7cm}p{0.7cm}p{1.3cm}p{1.3cm}}
\hline\noalign{\smallskip}
Framework & DAL & Prop. & Prop. & DAL & Prop.\\
\textit{target} & MC & MC & MC & Heterog. & Heterog. \\
\textit{mapping} & fixed & fixed & free & - & - \\
\hline 
Carrizo  & 5.5 & 7.1 & 8.8 & \textit{n/a} & 47.4\\
\hline 
i7 & 21.1 & 30.4 & 32.8 & \textit{n/a} & 83.8 \\
\hline
\noalign{\smallskip}
\end{tabular}
\end{table}

\subsection{Results}
\label{ssec:results}

Results for executing the Motion Detection and DPD applications on the proposed framework, and the reference framework DAL \cite{Schor12, Schor13} are presented in Table~\ref{table:motion} and Table~\ref{table:dpd}.

On both platforms the GPU-accelerated version of Motion Detection invoked a GPU driver issue on the DAL framework when the token rate was increased beyond 1. To enable benchmarking, the problem was circumvented by using 640x480 frames with token rate 1 instead of 320x240 frames with token rate 4 on GPU-accelerated DAL experiments. This alternative setting left the number of computations and OpenCL work dimensions identical, and hence the results are directly comparable to other results in Table~\ref{table:motion}.

For the Motion Detection application the proposed framework provided a performance advantage of 20\% compared to DAL on the 4 processors of the Carrizo platform, and there was no evident performance difference between fixed and free actor-to-core mappings. Using the proposed framework, on this platform the GPU provided a considerable 9.5$\times$ speedup compared to the best multicore-only results. Comparing the GPU-accelerated programs, the proposed approach was 58\% faster than DAL.

On the i7 platform multicore-only execution yielded 30\% higher throughput on the proposed framework than on DAL. When GPU-accelerated programs are compared, the proposed framework was 40\% faster than DAL.

For the DPD application the proposed framework yielded a 29\% speedup over DAL with fixed actor-to-core mapping on Carrizo multicore-only. The speedup gap increased to 60\% when the mapping was left to be decided by the underlying OS. On the proposed framework, the use of GPU acceleration provided a speedup of 5.4$\times$ compared to best multicore results, \textit{whereas on the DAL platform GPU acceleration was not possible as the DAL framework does not support dynamic data rate actors on the GPU.}

On the i7 platform the DPD application executed 44\% faster on the proposed framework (fixed mapping) than on DAL when only the multicore chip was used. The gap further increased to 55\% when the actor-to-core mapping was left to be decided by the OS. Finally, GPU acceleration enabled by the proposed framework yielded a 2.6$\times$ speedup compared to the best multicore results.

\section{Discussion and Future work}
\label{sec:discussion}

The results presented in Section~\ref{ssec:results} illustrate that the proposed framework
\begin{itemize}
  \item Provides 20\%-60\% higher throughput than the reference framework, and
  \item Enables executing applications with dynamic dataflow behavior on the GPU.
\end{itemize}

From the viewpoint of the application programmer, DAL and the proposed framework are to a great extent similar, differing mainly in the way actors are written for GPU targets. In DAL, it is in principle possible to write an actor in C language and have it executed by the framework either on GPPs or on the GPU without code modifications, however such actors are restricted to static data rates. At the moment the proposed approach requires an actor to be written in OpenCL C for GPU execution.

It is necessary to state that although the proposed approach overcomes DAL in terms of performance, DAL provides a number of features that are unavailable in the proposed framework, such as targeting distributed systems, error-resilience via spare core allocation and support for multiple simultaneous applications.

One of the limitations of the proposed framework is that an actor port may have at maximum two different data rates. The consequence of this restriction is that for applications such as Dynamic Predistortion, where the data path is arbitrarily changing at run time, the token rate must for the dynamic part of the network be kept at 1, otherwise the network has a risk of deadlocking. 

Hence, for future development of the framework the most clear direction is relaxation of token rate restrictions without sacrificing the efficiency of the framework. In the same vein, it is mandatory to present a formal definition of the framework's Model of Computation and analyze its properties to e.g. identify necessary conditions for avoiding the possibility of deadlock.

Another likely direction of future work is to make the API compatible with DAL. This would have two benefits: 1) applications from the DAL repository could directly be executed on the proposed framework, and 2) it would be possible to write programs for the proposed framework using the RVC-CAL dataflow language by means of the Open RVC-CAL Compiler DAL Backend \cite{Boutellier15S}.

\section{Conclusion}
\label{sec:conclusion}

We have presented a novel dataflow-flavored framework for efficient programming of heterogeneous multicore platforms. Compared to the state-of-the-art, the proposed framework pioneers in enabling the execution of dynamic dataflow actors on GPU devices.

Claims on the proposed framework's efficiency and features have been demonstrated with two applications, video motion detection and dynamic predistortion filtering. Experiments have shown that the proposed framework provides 20\% to 60\% higher throughput compared to the well-known DAL framework. Moreover, the proposed framework is capable of GPU-accelerating actors that have dynamic data rates, a feature that was measured to improve throughput up to 5$\times$.

\section{Acknowledgements}

This work was funded by Academy of Finland project UNICODE. 



\balance


\begin{thebibliography}{10}

\bibitem{Hyunh14}
H.~P. Huynh, A.~Hagiescu, O.~Z. Liang, W.-F. Wong, and R.~S.~M. Goh,
\newblock ``Mapping streaming applications onto {GPU} systems,''
\newblock {\em IEEE Transactions on Parallel and Distributed Systems}, vol. 25,
  no. 9, pp. 2374--2385, 2014.

\bibitem{Schor13}
L.~Schor, A.~Tretter, T.~Scherer, and L.~Thiele,
\newblock ``Exploiting the parallelism of heterogeneous systems using dataflow
  graphs on top of {OpenCL},''
\newblock in {\em IEEE Symposium on Embedded Systems for Real-time Multimedia
  (ESTIMedia)}, 2013, pp. 41--50.

\bibitem{Sbirlea12}
A.~Sb\^{\i}rlea, Y.~Zou, Z.~Budiml\'{\i}c, J.~Cong, and V.~Sarkar,
\newblock ``Mapping a data-flow programming model onto heterogeneous
  platforms,''
\newblock in {\em ACM SIGPLAN/SIGBED International Conference on Languages,
  Compilers, Tools and Theory for Embedded Systems (LCTES)}, 2012, pp. 61--70.

\bibitem{Kahn74}
G.~Kahn,
\newblock ``The semantics of a simple language for parallel programming,''
\newblock in {\em Information processing}, J.~L. Rosenfeld, Ed., Stockholm,
  Sweden, 1974, pp. 471--475, North Holland.

\bibitem{Lee87}
E.~A. Lee and D.~G. Messerschmitt,
\newblock ``Synchronous data flow,''
\newblock {\em Proceedings of the IEEE}, vol. 75, no. 9, pp. 1235--1245, 1987.

\bibitem{Lund15}
W.~Lund, S.~Kanur, J.~Ersfolk, L.~Tsiopoulos, J.~Lilius, J.~Haldin, and
  U.~Falk,
\newblock ``Execution of dataflow process networks on {OpenCL} platforms,''
\newblock in {\em Euromicro International Conference on Parallel, Distributed
  and Network-Based Processing (PDP)}, 2015, pp. 618--625.

\bibitem{Schor12}
L.~Schor, I.~Bacivarov, D.~Rai, H.~Yang, S.-H. Kang, and L.~Thiele,
\newblock ``Scenario-based design flow for mapping streaming applications onto
  on-chip many-core systems,''
\newblock in {\em International Conference on Compilers, Architectures and
  Synthesis for Embedded Systems (CASES)}, 2012, pp. 71--80.

\bibitem{Henzinger01}
T.~A. Henzinger, B.~Horowitz, and C.~M. Kirsch,
\newblock ``Giotto: A time-triggered language for embedded programming,''
\newblock in {\em Embedded Software: First International Workshop, ({EMSOFT})},
  T.~A. Henzinger and C.~M. Kirsch, Eds., 2001, pp. 166--184.

\bibitem{Bilsen96}
G.~Bilsen, M.~Engels, R.~Lauwereins, and J.~Peperstraete,
\newblock ``Cycle-static dataflow,''
\newblock {\em IEEE Transactions on Signal Processing}, vol. 44, no. 2, pp.
  397--408, Feb 1996.

\bibitem{Mattavelli10}
M.~Mattavelli, I.~Amer, and M.~Raulet,
\newblock ``The {Reconfigurable} {Video} {Coding} standard [standards in a
  nutshell],''
\newblock {\em IEEE Signal Processing Magazine}, vol. 27, no. 3, pp. 159--167,
  2010.

\bibitem{Berg08}
H.~Berg, C.~Brunelli, and U.~Lucking,
\newblock ``Analyzing models of computation for software defined radio
  applications,''
\newblock in {\em International Symposium on System-on-Chip}, 2008, pp. 1--4.

\bibitem{Buck93}
J.~T. Buck and E.~A. Lee,
\newblock ``Scheduling dynamic dataflow graphs with bounded memory using the
  token flow model,''
\newblock in {\em International Conference on Acoustics, Speech, and Signal
  Processing (ICASSP)}, April 1993, vol.~1, pp. 429--432 vol.1.

\bibitem{Plishker08}
W.~Plishker, N.~Sane, M.~Kiemb, K.~Anand, and S.~S. Bhattacharyya,
\newblock ``Functional {DIF} for rapid prototyping,''
\newblock in {\em IEEE/IFIP International Symposium on Rapid System Prototyping
  (RSP)}, 2008, pp. 17--23.

\bibitem{Lee95}
E.~A. Lee and T.~M. Parks,
\newblock ``Dataflow process networks,''
\newblock {\em Proceedings of the IEEE}, vol. 83, no. 5, pp. 773--801, 1995.

\bibitem{Lee09}
E.~A. Lee and E.~Matsikoudis,
\newblock ``The semantics of dataflow with firing,''
\newblock in {\em From Semantics to Computer Science: Essays in memory of
  Gilles Kahn}, G.~Huet, G.~Plotkin, J.J. L\'evy, and Y.~Bertot, Eds. Cambridge
  University Press, 2009.

\bibitem{Tretter15}
A.~Tretter, J.~Boutellier, J.~Guthrie, L.~Schor, and L.~Thiele,
\newblock ``Executing dataflow actors as {Kahn} processes,''
\newblock in {\em International Conference on Embedded Software (EMSOFT)},
  2015, pp. 105--114.

\bibitem{Boutellier15T}
J.~Boutellier, J.~Ersfolk, J.~Lilius, M.~Mattavelli, G.~Roquier, and
  O.~Silv\'en,
\newblock ``Actor merging for dataflow process networks,''
\newblock {\em IEEE Transactions on Signal Processing}, vol. 63, no. 10, pp.
  2496--2508, May 2015.

\bibitem{Boulos14}
V.~Boulos, S.~Huet, V.~Fristot, L.~Salvo, and D.~Houzet,
\newblock ``Efficient implementation of data flow graphs on multi-{GPU}
  clusters,''
\newblock {\em Journal of Real-Time Image Processing}, vol. 9, no. 1, pp.
  217--232, 2014.

\bibitem{Gautier13}
T.~Gautier, J.~V.~F. Lima, N.~Maillard, and B.~Raffin,
\newblock ``{XKaapi}: A runtime system for data-flow task programming on
  heterogeneous architectures,''
\newblock in {\em International Symposium on Parallel and Distributed
  Processing (IPDPS)}, 2013, pp. 1299--1308.

\bibitem{Thies02}
W.~Thies, M.~Karczmarek, and S.~Amarasinghe,
\newblock ``Streamit: A language for streaming applications,''
\newblock in {\em Compiler Construction}, R.~Nigel Horspool, Ed., vol. 2304 of
  {\em Lecture Notes in Computer Science}, pp. 179--196. Springer, 2002.

\bibitem{Boutellier15S}
J.~Boutellier and T.~Nyl\"anden,
\newblock ``Programming graphics processing units in the {RVC-CAL} dataflow
  language,''
\newblock in {\em IEEE Workshop on Signal Processing Systems (SiPS)}, 2015, pp.
  1--6.

\bibitem{Boutellier15G}
J.~Boutellier and A.~Ghazi,
\newblock ``Multicore execution of dynamic dataflow programs on the
  {Distributed} {Application} {Layer},''
\newblock in {\em IEEE Global Conference on Signal and Information Processing
  (GlobalSIP)}, 2015, pp. 893--897.

\end{thebibliography}
\end{document}